# Demystifying Weak Measurements*

R. E. Kastner**

March 22, 2017

ABSTRACT. A large literature has grown up around the proposed use of 'weak measurements' (i.e., unsharp measurements followed by post-selection) to allegedly provide information about hidden ontological features of quantum systems. This paper attempts to clarify the fact that 'weak measurements' involve strong (projective) measurements on one (pointer) member of an entangled system. The only thing 'weak' about such measurements is that the correlation established via the entanglement does not correspond to eigenstates of the 'weakly measured observable' for the remaining component system(s) subject to the weak measurement. All observed statistics are straightforwardly and easily predicted by standard quantum mechanics. Specifically, it is noted that measurement of the pointer steers the remaining degree(s) of freedom into new states with new statistical properties—constituting a non-trivial (even if generally small) disturbance. In addition, standard quantum mechanics readily allows us to conditionalize on a final state if we choose, so the 'post-selection' that features prominently in time-symmetric formulations is also equipment from standard quantum theory. Assertions in the literature that weak measurements leave a system negligibly disturbed, and/or that standard quantum theory is cumbersome for computing the predicted measurement results, are therefore unsupportable, and ontological claims based on such assertions need to be critically reassessed.

1. What are "weak measurements"?

There are two ways in which the term "weak measurement" is used in the literature:
 i.   an unsharp or non-ideal measurement
 ii.  an unsharp measurement followed by a sharp post-selection measurement

Case (i) has been studied and quantified extensively by Busch et al (1995) and many others. Case (ii), which adds the post-selection measurement to (i), is used primarily by proponents of the Two-State Vector Formalism (TSVF), first proposed by Aharonov and Vaidman (1990), to argue for unobservable but real hidden ontological properties of the measured system. Claims about surprising and/or retrocausal properties of quantum systems based on (ii) have appeared, for example, in Aharonov, Y., Albert, D.Z., Vaidman, L. (1988) and Aharonov et al (2002). The specific claims addressed in this paper are contained in Aharonov, Cohen and Elitzur (2015), henceforth "ACE". That paper considered (in part) a two-particle system, the weakly measured system $S$ and a pointer $P$; this two-particle system is what is discussed herein. Of course, there are more complicated experiments with more degrees of freedom, but the basic principles are the same (i.e., that disturbances taken into account via standard quantum mechanics readily account for all the observed phenomena).

Regarding (i), in addition to the pioneering and comprehensive studies of Busch et al, a clear pedagogical presentation is given in Bub (1997), Section 5.3. The defining characteristic of an unsharp measurement is that the pointer system (call it P) is entangled with the target system (call it S) in such a way that there is a nonvanishing amplitude for a "wrong answer" regarding the measured S observable $O$. That is, there are error terms, such that when P is detected, S does not end up in an eigenstate of $O$. This means that when P gives a pointer reading corresponding to a particular value of $O$, the system S is not in the corresponding eigenstate of $O$, and it cannot be inferred that S has that property; at least not on the basis of that measurement result.

Nevertheless, when P is detected (which is a crucial part of the weak measurement), both systems are collapsed into pure states, and their entanglement is destroyed. The latter point seems to be overlooked when claims are made based on the use of (ii), i.e. when post-selection is added to the process and so-called "weak values" are considered as having ontological applicability to individual systems. The "weak value" of an observable $O$, $O_W$, is defined in terms of pre- and post-selection states $|x\rangle$ and $|y\rangle$ as $O_W := \frac{\langle y|O|x\rangle}{\langle y|x\rangle}$. Thus, it is simply a normalized transition amplitude (matrix element) for $O$. It is important to note that this is a theoretical quantity defined in terms of operators and states, without regard to any particular process of measurement. Thus, the term 'weak value' is something of a misnomer: there is nothing 'weak' about the value itself.

Of course, one can use unsharp measurements over many experimental runs to obtain a statistical measure of the weak value, but it would be a mistake to conclude on that basis that the term 'weak value' is only defined or definable in the context of unsharp measurements. One can always compute the normalized matrix element for any operator O and any pre- and post-selection states, and get a well-defined result (of course, in general it will be complex, since it's really just an amplitude). So it is a theoretically well-defined quantity independently of any specified measurement process.

In any case, if one is 'measuring the weak value' via unsharp measurements, it is crucial to note that in any particular run, the pointer will be detected in a random state of the pointer basis (with a probability given by the Born Rule). The pointer will not, in general, be detected at (or even necessarily near) the weak value in any individual run; the 'measurement of the weak value' is a determination of the *average* pointer outcome, found over many runs.[1] Thus, for clarity, it should be noted that in this paper, the term 'weak value' means its exact definition above, not any particular method of 'measuring the weak value,' nor any putative 'weak element of reality,' as discussed in Vaidman (1996). In fact the existence of the latter is an aspect of what is under dispute here.

A variant of (ii) is to delay the detection of P until after S has been post-selected; then the post-selection of S collapses P. So either way, in the case of the two entangled

---

[1] In this procedure, the 'weak value' is no more an ontological property of any individual system/pointer pair than an average height is for each member of a large group of people, although of course their heights collectively contribute to the average.

systems discussed in ACE, complete disentanglement takes place. We consider both situations below.

2. Pointer measured first

Let us begin with the case in which the pointer particle P is detected before the target system S is post-selected--which is used, for example, in ACE to argue for the TSVF ontology. For clarity and specificity, consider a particular simple example. We will call the experimenters Alice and Bob to make contact with the argument in ACE. An electron is prepared in the state 'up along x', $|x+\rangle$; Alice performs a a weak (unsharp) measurement of its spin along Z (using P), and finally, Bob performs a sharp measurement of the Z-spin of S, yielding either $|z+\rangle$ or $|z-\rangle$. Thus we have the following steps:

I. An electron S in state $|x+\rangle = \frac{1}{\sqrt{2}}[|z+\rangle + |z-\rangle]$ is emitted toward a Stern-Gerlach apparatus equipped with a weakly measuring photon gun emitting photons whose wavelength approaches the dimensions of the experimental apparatus (so that we are in the realm of wave optics rather than ray optics, leading to an imprecise measurement). The electron plays the part of S and the photon plays the part of P. The photon has an error amplitude $b$ of being scattered into to the wrong detector ; so, for example, a state $|z+\rangle$ will result in photon detection at the correct 'up' detector with an amplitude of $a$, and will result in photon detection at the wrong 'down' detector with an amplitude of $b<a$. ($b^2 + a^2 = 1$.)

Specifically, the unsharp measurement interaction gives rise to the evolution:

$$|P_0\rangle|z+\rangle \rightarrow |P+\rangle|z+\rangle = [a|u\rangle|z+\rangle + b|d\rangle|z+\rangle] \qquad (1a)$$

$$|P_0\rangle|z-\rangle \rightarrow |P-\rangle|z-\rangle = [a|d\rangle|z-\rangle + b|u\rangle|z-\rangle] \qquad (1b)$$

where $|P_0\rangle$ is the 'ready' state of the photon serving as the pointer, and |u> and |d> are the orthogonal 'up' or 'down' pointer basis states.[2] Note however that $|P+\rangle$ and $|P-\rangle$ are not orthogonal. This is the distinguishing feature of an 'unsharp' measurement.

For the electron S prepared in the state $|x+\rangle = \frac{1}{\sqrt{2}}[|z+\rangle + |z-\rangle]$, after the interaction an entangled two-particle state has been created:

$$|\Psi\rangle = \frac{1}{\sqrt{2}}[a|u\rangle|z+\rangle + a|d\rangle|z-\rangle + b|u\rangle|z-\rangle + b|d\rangle|z+\rangle] \qquad (2)$$

At this point, neither S nor P is in a pure state; each is in an improper mixed state.[3] Note that the amplitudes $a$ describe states with the correct correspondence between S and P, while the amplitudes $b$ describe states with incorrect correspondence. In a sharp measurement, $a=1$ and $b=0$, and we only have the 'correct' terms.

II. Alice records the photon (P) detections. When Alice detects the photon P, the composite state becomes disentangled, as follows. If the photon is detected as 'up', |u>, by Alice, the total system is projected into the attenuated, disentangled two-particle state

$$|\Psi+\rangle = \frac{1}{\sqrt{2}}[a|u\rangle|z+\rangle + b|u\rangle|z-\rangle] = \frac{1}{\sqrt{2}}|u\rangle[a|z+\rangle + b|z-\rangle] \qquad (3)$$

I.e., this is now a product state of the component systems P and S. Factoring out the P state $\frac{1}{\sqrt{2}}|u\rangle$, we see that the electron S is now in an "unsharp" pure state of

$$|e_u\rangle = [a|z+\rangle + b|z-\rangle] \qquad (4)$$

---

[2] ACE use a continuous pointer basis, which simply means that there are infinitely many possible 'error' states for the pointer, within a given range. For the continuous pointer position basis $\{X\}$, the entanglement would be set up through the evolution $|P_0\rangle|z+\rangle \to \int dX \langle X|P+\rangle|X\rangle|z+\rangle$ and $|P_0\rangle|z-\rangle \to \int dX \langle X|P-\rangle|X\rangle|z-\rangle$. P is always detected in some position state $|X\rangle$ which, due to the overlap of $|P+\rangle$ and $|P-\rangle$, cannot be clearly identified with either 'up' or 'down' on the part of the measured system. We use a discrete two-valued pointer basis here, $\{|u\rangle, |d\rangle\}$, to clarify the basic concepts involved.

[3] An improper mixed state for a component system is obtained by tracing out the degrees of freedom of its entangled partner. In general, it cannot be interpreted as a statistical mixture over unknown but ontologically applicable pure states of the system, hence the term 'improper'.

This is no different conceptually from an EPR-type experiment in which the detection of one of the electrons projects the other into a pure state.

Thus, over the entire "weak measurement" process, the electron's (S's) state has undergone the following changes:

1. from a pure prepared state (|x+>)
2. to an improper mixed state (as a subsystem of the entangled state (1))
3. to an entirely different pure state (3)

Clearly, therefore (unless we have *exactly a=b=$\frac{1}{\sqrt{2}}$*), S has been non-negligibly disturbed. The only difference between this 'weak measurement' and the standard 'strong' or 'sharp' measurement is that S is not in an eigenstate of its observable.

For purposes below, note that S's unsharp (but still pure!) state can be written in the x-spin basis as

$$|e_u\rangle = \frac{1}{\sqrt{2}}[(a+b)|x+\rangle + (a-b)|x-\rangle] \qquad (5)$$

So that, if the Z measurement were a sharp one (a=1; b=0), $|e_u\rangle$ would be the state $\frac{1}{\sqrt{2}}[|x+\rangle + |x-\rangle] = |z+\rangle$.

On the other hand, if the *z* measurement were maximally weak (a=b=$\frac{1}{\sqrt{2}}$), $|e_u\rangle$ would be the state |x+>; i.e., we would get back the original prepared state. This is the only case in which S can be said to be undisturbed.

Meanwhile, a detection of P as 'down' by Alice projects the electron into the unsharp pure state

$$|e_d\rangle = [a|z-\rangle + b|z+\rangle] \qquad (6)$$

Again, this can be written in the x-spin basis as

$$|e_d\rangle = \tfrac{1}{\sqrt{2}}[(a+b)|x+\rangle - (a-b)|x-\rangle] \qquad (7)$$

So that, if the Z measurement were a sharp one (a=1; b=0), |e_d> would be |z-> ;
if Z measurement is maximally weak (a=b=1/√2) we again get back the original state, |x+>. In general, however, the two states |e_u> , |e_d> of S are distinct but not orthogonal, which is the distinguishing feature of a 'weak' or 'unsharp' measurement.

III. Now for the post-selection. Another experimenter, Bob, conducts a strong (sharp) measurement of the electron's spin along Z. Each electron reaches Bob in a particular unsharp pure state brought about by Alice's photon measurement: either |e_u> or |e_d>, each with a probability of ½. (At this stage, Alice knows this state but Bob does not.) If Bob's unknown state is |e_u> , the outcome z+ will occur with the probability *a*a*. If Bob's unknown state is |e_d >, the outcome z+ will occur with the probability *b*b*.  So Bob's probability of seeing 'z+' is  ½ *(a*a + b*b)* = ½ ; and similarly for the result 'down'.

    Alice's weak measurement result – the value obtained for P – will not always coincide with Bob's result for S because (for b =/= 0) her detections always leave an error component in the electron state. That is, Alice may find the photon in the 'up' detector; this means that the electron has been projected into the state e_u. But given that state, there is a probability of *b*b* that the electron will be detected at Bob's detector for |z->.

    It's important to keep in mind that Alice is actually measuring P (i.e., she is detecting P in eigenstates of the P observable, |u> or |d> ).  She is NOT measuring S, despite the fact that this process is called a 'weak measurement of S.' Now, if Alice and Bob apply a serial number to each electron/photon pair, and get together afterwards to compare notes, we will see something like the following (an X is placed next to Bob's 'wrong' sharp S outcomes based on the P outcome obtained by Alice):

| # | P(Alice) | S(Bob) |
|---|---|---|
| 1 | d | – |
| 2 | u | + |
| 3 | u | – (X) |
| 4 | d | – |
| 5 | u | + |
| 6 | d | + (X) |
| 7 | d | – |
| 8 | u | + |

Table 1

Depending on the values of the amplitudes *a* and *b*, the outcomes for P and S will line up more or less accurately. (This distribution has $b^2 = 1/4$, to illustrate the effect.) Bob's post-selection outcome (indicated in the right-hand column) creates sub-ensembles, i.e., two induced distributions over Alice's P outcomes:

| # | P(Alice) | S(Bob) |
|---|---|---|
| 2 | u | + |
| 5 | u | + |
| 6 | d | + (X) |
| 8 | u | + |

Table 2a

| # | P(Alice) | S(Bob) |
|---|---|---|
| 1 | d | – |
| 3 | u | – (X) |
| 4 | d | – |
| 7 | d | – |

Table 2b

We see that there is a correlation (within error bars of $b^2 = 1/4$) between Alice's result for P and Bob's result for S. Is this surprising? Does it show that the electron (within error bars) "knew ahead of time" what Bob's result would be? Let's explicitly put in the unsharp but pure state obtaining for S as a result of each of Alice's P outcomes:

| # (S:+) | P(Alice); S unsharp state | S(Bob) |
|---|---|---|
| 2 | u; $|e_u\rangle = [a|z+\rangle + b|z-\rangle]$ | + |
| 5 | u; $|e_u\rangle = [a|z+\rangle + b|z-\rangle]$ | + |
| 6 | d; $|e_d\rangle = [b|z+\rangle + a|z-\rangle]$ | + (X) |
| 8 | u; $|e_u\rangle = [a|z+\rangle + b|z-\rangle]$ | + |

and

| #(S: -) | P(Alice); S unsharp state | S(Bob) |
|---|---|---|
| 1 | d; $|e_d\rangle = [b|z+\rangle + a|z-\rangle]$ | – |
| 3 | u; $|e_u\rangle = [a|z+\rangle + b|z-\rangle]$ | – (X) |
| 4 | d; $|e_d\rangle = [b|z+\rangle + a|z-\rangle]$ | – |
| 7 | d; $|e_d\rangle = [b|z+\rangle + a|z-\rangle]$ | – |

Tables 3a (top) and 3b (bottom)

These sub-ensembles show the following: when Bob finds S (the electron) in the state 'up', it is more likely (quantified by $|a|^2$) to have come from the state $|e_u\rangle$ than from $|e_d\rangle$; and when he finds S in the state 'down', it is more likely to have come from the unsharp state $|e_d\rangle$ than from $|e_u\rangle$. Recall that before Alice measured the pointer particle P, the electron S was prepared in the state $|x+\rangle$, which has no preference for either $|z+\rangle$ or $|z-\rangle$. But more generally, S could have been prepared in a state with more or less statistical tendency for $|z+\rangle$ or $|z-\rangle$, and then Alice's P outcomes would reflect that tendency, while Bob's S outcomes would reflect the statistical tendency of the unsharp S state that reaches him. These statistical tendencies, reflected in the data, are attributes of the state itself, and as such they do not require any hidden ontology. No matter how many such measurements one performs, if one looks only at the runs in which S was post-selected in a particular state $|i\rangle$ (as in the Tables 3a and 3b), those $S_i$ will always be more likely to have come from a state that statistically favored $|i\rangle$. And that greater likelihood will be correctly quantified by the quantum states themselves, not from any hidden ontology. That is, standard quantum theory based only on quantum states (without any hidden ontological property) perfectly well accounts for the correlations.

The crucial point here is that the electron S is already in a pure state, either $|e_u\rangle$ or $|e_d\rangle$, when Bob measures it. Disentanglement of S from P is complete upon Alice's measurement of P. When Alice performs repeated weak measurements on the same electron having serial number $S_i$ (including those for noncommuting observables), each such additional measurement $j$ of the same electron $S_i$ must use a new pointer system $P_{ij}$, and must create a new entanglement with $S_i$, based on whatever pure state $S_i$ ended up in after the last detection of the photon with serial number $P_{i,j-1}$. And every time a photon $P_{ij}$ is detected, $S_i$'s state is projected into a new pure state $|e\rangle_{ij}$, whose statistical properties are correctly predicted by standard quantum mechanics and will be duly reflected in the data gathered.

Thus the process of repeated weak measurements is a process of repeated collapses/disentanglements, re-entanglements of a particular electron $S_i$ in its new pure state with a fresh pointer $P_{ij}$, and more collapses/disentanglements with each detection of a new $P_{ij}$. This point, obscured in the ACE presentation, is important because it concerns the applicability of the data to the individual systems involved, which has a direct bearing on matters of ontology. Certainly, for example, one could not say that all these weak measurements apply to the same serial number understood as describing the initial electron/photon pair. With every additional measurement of the same electron $S_i$, Alice is using a new $P_{ij}$. And each of the strong measurements of $P_{ij}$ steers its entangled partner $S_i$ into a new collapsed pure state $|e\rangle_{ij}$, just as in an EPR situation. That new output pure state will of course reflect all the standard quantum statistics of the input pure state $|e\rangle_{i,j-1}$ resulting from the previous weak measurement.

Thus, the above process can hardly be described as leaving a system $S_i$ undisturbed or negligibly disturbed, as is often claimed in the literature on weak measurements. A series of $N$ projective measurements $j=\{1,N\}$ on entangled partners $P_{ij}$ of a system $S_i$ results in many nontrivial (even if small in magnitude) state transformations of $S_i$, all of which lead directly to the observed correlations, exactly as described by standard quantum mechanics.

## 3. S post-selected first

What if Bob were to perform his measurement before Alice detected P? Then Bob would be measuring S directly from the entangled state (2), i.e.:

$$|\Psi\rangle = \frac{1}{\sqrt{2}}[a|u\rangle|z+\rangle + a|d\rangle|z-\rangle + b|u\rangle|z-\rangle + b|d\rangle|z+\rangle] \tag{2}$$

When Bob finds the outcome z+ or z–, its entangled partner P is projected into one of the pure unsharp states (respectively):

$$|P+\rangle = [a|u\rangle + b|d\rangle] \tag{8a}$$

or

$$|P-\rangle = [a|d\rangle + b|u\rangle] \tag{8b}$$

These are analogous to the states $|e_u\rangle$ and $|e_d\rangle$ for S, eqs. (4) and (6) above; they reflect the basic unsharp measurement evolution (see eqs (1) ). Thus, we end up with the same situation as before, but with the roles of S and P reversed: in a sense, S 'weakly measures' P. The entanglement is completely destroyed by S's post-selection and Alice receives an unsharp pure state of P, which she then collapses either to |u> or |d> upon detecting the photon P. The pattern of correlations has the same origin, and is exactly as predicted by standard quantum theory.

ACE also consider an EPR-correlated pair subject to a series of weak measurements, followed by strong measurements; this situation will be addressed in a separate work, but the basic principles are the same: standard quantum theory transparently predicts all the observed correlations, once post-selection is taken into account. It should also be clarified that taking post-selection into account does not indicate any departure from standard one-vector quantum theory, as ACE suggest. For example, they say:

```
     "The mathematical simplicity and conceptual elegancy
[sic] of the TSVF are evident. When the pre- and post-
selected states are set, we immediately know what would be
```

```
all the weak values of this sub-ensemble, without the need
of calculating the consecutive non-commuting biases. "
``` (ACE, p. 15, preprint version)

However, the calculational simplicity referred to is just as much equipment from the standard formalism as it is from the TSVF, since 'weak values' are just operator matrix elements normalized by the overlap of the pre- and post-selection states. That is, there is no sense in which a 'weak value' is uniquely a theoretical object of TSVF as opposed to the standard quantum theory. Yes, a normalized transition amplitude was *arrived at* by researchers exploring post-selection and time-symmetry issues, but here we have to be wary of the genetic fallacy: the idea that the content of a concept or proposition necessarily has something to do with the way in which it was conceived.[4]

Post-selection simply defines a set of sub-ensembles, as illustrated in the above examples. The fact that one can create sub-ensembles defined by a post-selection outcome and find statistical correlates of operator matrix elements for those sub-ensembles is being portrayed as evidence for a particular ontology in which the post-selection is viewed as ontologically symmetric with the preselection, and the elements of each sub-ensemble were predestined to be so selected. But it is not at all clear that these phenomena really warrant such an inference. On the contrary, taking post-selection into account does not appear to have bearing, one way or the other, on such ontological questions, since all the calculations (including the simple, elegant ones using the operator matrix elements) involve theoretical elements of standard quantum mechanics.

5. Conclusion

We have looked at some examples of weak and strong measurements on a single particle along the lines discussed by ACE (2015), and found that all correlations seen from the vantage point of post-selected sub-ensembles are transparently accounted for in standard quantum mechanics, based on the quantum states resulting from the weak (unsharp) and strong (sharp) measurements. That is, the correlations arise directly from the quantum states resulting from each of the measurements (whether weak or strong), rather than from any additional ontological properties characterized by 'weak values' (which are just operator matrix elements normalized by a post-selection outcome).

---

[4] For example, the Born Rule for the probabilities of measurement outcomes was arrived at via an educated guess by Max Born. Clearly this does not indicate that the Rule's ontological significance has something to do with guessing.

In view of the above, one might ask: can the putative existence of a hidden property (i.e. the spin of S in this case) be said to be corroborated by a given procedure, if assuming the existence of that property has no predictive consequences (or even theoretical correlates) that differ from those based on assuming its lack of existence? The answer would appear to be negative. Of course, there are other considerations that ACE (2015) being to bear on the interpretation of the measurement statistics they discuss; these will be addressed in a separate work.[5] The main point of this paper is simply to make clear that a 'weak measurement' of a given system S is actually a strong (sharp) measurement of its entangled partner P, which always completely destroys their entanglement; and that standard quantum mechanics gives a simple and straightforward account of all the observed correlations. Thus, the claim that standard quantum theory is somehow awkward or inelegant in accounting for the correlations seems markedly overstated.

Perhaps the simplest way to see what is arguably an inferential fallacy underlying the ontological claims surrounding 'weak measurement' statistics is through the following consideration. Standard QM describes a branching probability structure—that is, a given initial state leads to many possible final outcomes, branching at each measurement node. In contrast, TSVF assumes no branching--a system is taken as having determinate initial and final states (i.e., the "pre- and post-selection"), and the only uncertainty regards the outcome of the 'weakly measured' observable in between those initial and final states. But one gets exactly the same statistics for the 'weakly measured observable', just as easily in standard quantum theory, when one conditionalizes on a particular final outcome. Thus, the statistics underdetermine the ontology. Nature could have the branching structure, or she could have the non-branching structure. The calculations have nothing to say about which is the case, since they are exactly the same.

---

[5] Such considerations include the ambiguity of time order of the measurements on spacelike-separated systems. This issue raises interesting interpretive questions and will be considered in light of other time-symmetric interpretations in a separate work. But it does not in itself single out TSVF as the appropriate ontology. Also, the present author is aware that much more complicated experiments can and have been carried out, with entanglement of greater than two degrees of freedom. This does not affect the basic argument, which is that the behavior all such systems is exactly as predicted by standard quantum mechanics, and that one can always conditionalize on any final state of interest to make use of the more 'elegant' calculations advertised as a putative advantage of the TSVF formalism.

Furthermore, standard quantum mechanics cannot be faulted for 'having to conditionalize on a final outcome': TSVF does exactly that while presupposing, without independent empirical warrant, that there was never any branching structure in the first place. By illustration, it may seem 'simpler' or 'more elegant' to conclude, for example, that M. Brunet was retroactively made French *because* he was found to have arrived on a plane from France (the plane arrival playing the part of the 'post-selection'), but that does not mean that it is the correct account of why M. Brunet is in fact French. It's just statistically more likely for people arriving on planes from France to be French. So this is a matter of taking care with what sorts of causal/ontological inferences can be justified based on observed statistical correlations. It has nothing to do with whether the systems are classical or quantum in nature—the latter simply changes what sorts of statistics one gets.

Finally, it should be noted that this paper should not be interpreted as an effort to shield 'standard quantum mechanics' from reformulation or reinterpretation. The question is, what sorts of arguments are effective in supporting alternative formulations? This paper attempts to highlight what appears to be a basic inaccuracy in arguments that TSVF reflects a particular sort of hidden ontology: namely, the claim that disturbances to the system due to weak measurements can be neglected; and also to emphasize that the claimed need for TSVF in order to account for the correlations seems overstated, given that (in light of the analysis presented herein) they appear to be perfectly transparent and unproblematic under standard quantum mechanics.

Acknowledgments

The author is grateful to Avshalom Elitzur and Eliahu Cohen for interesting and valuable discussions and correspondence.

References

Aharonov, Y., Cohen, E., Elitzur, A. (2015). *Ann. Phys. 355* 258-268. (Preprint: http://arxiv.org/abs/1206.6224)

Aharonov, Y., Albert, D.Z., Vaidman, L. (1988) Physical Review Letters. **60** (14): 1351–1354.

Aharonov, Y. and Vaidman, L. (1990). *Phys. Rev. A 41*, 11.

Aharonov, Y., Alonso Botero, Sandu Popescu, Benni Reznik, Jeff Tollaksen (2002). Revisiting Hardy's paradox: counterfactual statements, real measurements, entanglement and weak values. Phys. Lett. A, 301: 130–8.

Bub, J. (1997). *Interpreting the Quantum World*. Cambridge: Cambridge University Press.

Busch, P., Grabowski, M., Lahti, P. (1996*). Operational Quantum Physics*. Berlin, Heidelberg: Springer-Verlag.

Vaidman. L. (1996). 'Weak-Measurement Elements of Reality, ' *Foundations of Physics 26* , 895-90.